\documentclass[twoside,leqno,twocolumn]{article}

\usepackage[letterpaper]{geometry}

\usepackage{ltexpprt}

\usepackage[utf8]{inputenc} 
\usepackage{hyperref}       
\usepackage{url}            
\usepackage{booktabs}       
\usepackage{amsfonts}       
\usepackage{nicefrac}       
\usepackage{microtype}      

\usepackage{xcolor}
\usepackage{graphicx}
\usepackage{placeins}
\usepackage{float}
\usepackage{array}
\usepackage{caption}
\usepackage{subcaption}

\usepackage{amsmath}
\usepackage{amssymb}

\def\norm#1{\left\| #1 \right\|}

\newcommand\Tstrut{\rule{0pt}{2.6ex}}         
\newcommand\Bstrut{\rule[-0.9ex]{0pt}{0pt}}   
\begin{document}
\title{A Study of Performance of Optimal Transport
}
\author{Yihe Dong\thanks{Microsoft Research.}
\and Yu Gao\thanks{Georgia Tech. Part of this work was done while at Microsoft Research.}
\and Richard Peng \footnotemark[2]
\and Ilya Razenshteyn\footnotemark[1]
\and Saurabh Sawlani\thanks{Georgia Tech.}}

\date{}

\maketitle



\begin{abstract} \small\baselineskip=9pt
We investigate the problem of efficiently computing optimal transport (OT) distances, which is equivalent to the node-capacitated minimum cost maximum flow problem in a bipartite graph. We compare runtimes in computing OT distances on data from several domains, such as synthetic data of geometric shapes, embeddings of tokens in documents, and pixels in images. We show that in practice, combinatorial methods such as network simplex and augmenting path based algorithms can consistently outperform numerical matrix-scaling based methods such as Sinkhorn \cite{Cuturi13} and Greenkhorn \cite{AltschulerWR17}, even in low accuracy regimes, with up to orders of magnitude speedups. Lastly, we present a new combinatorial algorithm that improves upon the classical Kuhn-Munkres algorithm. 
\end{abstract}

\section{Introduction}

The optimal transport (OT) problem can be defined formally as follows:
given a supply distribution $r\in \mathbb{R}^n$ and a demand distribution $c \in \mathbb{R}^m$, find the optimal flow $X\in \mathbb{R}^{n\times m}$ that minimizes the overall transportation cost: 
\begin{align} \label{eq:statement}
    \min_X \langle C, X\rangle, \text{ s.t. } X \mathbf{1}_r = r \text{ and } X^T \mathbf{1}_c = c,
\end{align}
where $C$ denotes the cost matrix, $X$ the flow matrix, and $\mathbf{1}_r \in \mathbb{R}^m$ and $\mathbf{1}_c \in \mathbb{R}^n$ the all ones vectors.
The Hitchcock-Koopmans transportation problem is a special case of the above -- where $r$ and $c$ consist of only positive integers. Further, the case when $r$ and $c$ are all-ones vectors is known as the assignment problem.

The OT problem has recently received much attention as an algorithmic subroutine,
due to its relevance to the machine learning
community \cite{Cuturi13, BenamouCCNP15, GenevayCPB16, AltschulerWR17, DvurechenskyGK18, BlanchetJKS18, LinHJ19, Quanrud19}.

At first glance, Equation~\ref{eq:statement} can simply be solved exactly using an efficient linear program solver \cite{LeeS14} in about $n^{2.5}$ time.
However, to our knowledge, there have yet to be publically available packages that implement these more numerically driven methods efficiently.
Numerical techniques based on entropic regularization \cite{Cuturi13, BenamouCCNP15, GenevayCPB16, AltschulerWR17} have been shown to be scalable, both in terms of practice and their asymptotic runtimes.
In particular, \cite{Cuturi13} and \cite{AltschulerWR17} are near-linear time additive approximation algorithms for the optimal transport problem.

On the other hand, many combinatorial techniques also give exact algorithms for the OT problem.
The Kuhn-Munkres (a.k.a. Hungarian) algorithm \cite{Kuhn55, Kuhn56, Munkres57} solves the assignment problem exactly in $O(n^3)$ time.
For the optimal transport problem, Gabow and Tarjan \cite{GabowT91} gave an $O(n^{2.5}\log (n N))$ time
algorithm, involving scaling costs to integers and solving the transportation problem.
Here, $N$ refers to the largest element in the cost matrix.
Using the technique of scaling to integer demands and supplies, min-cost flow algorithms such as network simplex \cite{AhujaMO93} also provide exact algorithms for the optimal transport problem in $O(n^3 \log n \log (nN))$ time. Recently, Lahn \textit{et. al.} \cite{LahnMR19} modified the Gabow-Tarjan algorithm to give a faster approximation algorithm.

The computation of optimal transport distances has recently gained momentum across multiple problem domains, such as natural language processing \cite{KusnerSKW15} and image retrieval \cite{RubnerTG00, Cuturi13, SandlerL11}.
Additionally, computing optimal transport distances between probability distributions
has direct applicability to tasks involving unsupervised learning \cite{ArjovskyCB17, BigotGKL17}, semi-supervised learning \cite{SolomonRGB14}, clustering \cite{HoNYBHP17}, and several other applications \cite{KolouriPTSR17}.
Therefore, it is essential to develop a good understanding of the efficiency and effectiveness of the various methods across different types and sizes of data.
As far as we are aware, there have not been comprehensive studies that rigorously cross-examine the accuracies and performances of the classical and more recent methods to solve the OT problem. We seek to fill that void in this study, for a variety of relevant datasets. We also present a new batched KM algorithm that significantly outperforms classical KM in practice. 

In this paper, we compare various exact and approximate OT algorithms ranging from combinatorial methods to numerical ones based on matrix rescaling.
We perform comparisions in BLAS-optimized verisons of C++ when possible \footnote{Code available: \url{https://github.com/twistedcubic/fast_ot}}.

\section{Optimal Transport Algorithms}
\label{sec:algos}
In this section, we outline the efficient OT algorithms evaluated in our experiments.

\subsection{Network Simplex.}
Network simplex \cite{AhujaMO93} is a specialized version of the simplex algorithm that solves the min-cost max-flow problem on graphs. Formulating the min-cost max-flow problem as a linear program, each edge corresponds to a variable.
Similar to the set of basic variables in the original simplex algorithm, the network simplex algorithm maintains a spanning tree of edges. 

The following observation explains why it is sufficient to maintain a spanning tree.
Consider some flow on a graph. We say an edge is nonbasic if the amount of flow along it is not equal to $0$ nor its capacity. If there exists an undirected cycle formed by nonbasic edges, we can send flow along the cycle clockwise or counterclockwise to decrease the cost until some edge becomes basic. Thus, we observe that in an optimal solution, the nonbasic edges always form a spanning tree (or forest). 

The network simplex algorithm restricts its search amongst such spanning trees. In each step, the algorithm finds a basic edge (the entering edge) and the path in the spanning tree connecting the two endpoints of that edge. It then tries to push flow along the cycle formed by the edge and the path in a direction that decreases the total cost.
The entering edge generally becomes nonbasic and is then included in the spanning tree, while another edge in the tree path (the leaving edge) becomes basic and leaves the tree. 

\subsection{Matrix Scaling.}
Matrix scaling methods are developed to derive approximate solutions to the optimal transport linear problem. Here we discuss Sinkhorn \cite{Cuturi13} and Greenkhorn \cite{AltschulerWR17}.

To improve numerical stability of matrix scaling methods, we track and scale the logarithm of the flow values to avoid under- and overflows.

\subsubsection{Sinkhorn.}

Cuturi \cite{Cuturi13} opened up a new line of work that uses matrix scaling to find approximate solutions to the optimal transport problem, by finding the Sinkhorn distance $d_C^{\eta}(r, c):=\langle X^\eta, C\rangle$, with $X^{\eta}$ a solution to the OT problem with entropy regularization:
\[X^{\eta} = \text{argmin}_X \left(\langle X, C\rangle - \frac{1}{\eta} h(X)\right)\]
where $\eta > 0$, $h(X)=-\sum_{i,j} X_{i,j} \log(X_{i, j})$ denotes entropy, and the flow $X$ is optimized over the feasible region where $X \mathbf{1}_r = r$ and $X^T \mathbf{1}_c = c$.
The idea of entropic regularization for this problem, although not a novel one,
proves very effective in finding approximate solutions to the OT problem.

Adding the entropy term to the objective function enforces a structure to the solution $X^{\eta}$: $X^{\eta} = D_1 e^{-\eta C} D_2$, where $e^{-\eta C} \in \mathbb{R}^{n \times m}$ is the entrywise exponential of $-\eta C$, and $D_1 \in \mathbb{R}^{n\times n}$
and $D_2 \in \mathbb{R}^{m\times m}$ are diagonal matrices. 

Therefore, $X^{\eta}$ and hence the Sinkhorn distance $d_k$ can be computed by iterative matrix scaling on the matrix $e^{-\eta C} $, whereby all rows and all columns are alternately scaled to minimize the residue, i.e. the difference between current row or column sums and the target row or column sums.

After the residue $\norm{r-X \mathbf{1}_r}_1+\norm{c-X^T \mathbf{1}_c}_1$ drops below a preset threshold $\epsilon$, the flow $X$ is rounded so that the solution lies in the feasible region, i.e. $X \mathbf{1}_r = r$ and $X^T \mathbf{1}_c = c$.
As decribed in Section~\ref{sec-rounding}, this rounding does not necessasrily bring the solution closer to the optimum.


\subsubsection{Greenkhorn.}
Greenkhorn \cite{AltschulerWR17} is a greedy adaptation of Sinkhorn; instead of scaling all rows and all columns simultaneously, Greenkhorn scales one row or one column where the gain is the most, i.e. where the discrepancy between the current row or column sum and the target row or column sum is the largest. 
Greenkhorn brings the theoretical runtime down to $O(n^2\log(n)(N/\delta)^3)$.

\cite{AltschulerWR17} shows that Greenkhorn converges with fewer row/column updates than Sinkhorn. However, one Greenkhorn update is more expensive than one Sinkhorn update, as Sinkhorn updates all rows or all columns simultaneously, whereas Greenkhorn needs to keep track of which row and which column to update next.

\subsubsection*{Rounding.}
\label{sec-rounding}
Both Sinkhorn and Greenkhorn output not only an approximate OT distance, but also a feasible flow. This is achieved by a supplementary \emph{rounding} procedure \cite{AltschulerWR17}, which first scales all routed flows down to at most the target supplies and demands, and then distributes the leftover supplies proportional to the leftover demands. This routing of the leftover supplies, while ensuring feasibility of the solution, often results in a worse OT distance approximation on the datasets tested.

The effect of rounding on the objective can be bounded as follows: without rounding, since $$\norm{r-X \mathbf{1}_r}_1+\norm{c-X^T \mathbf{1}_c}_1\le \epsilon,$$ there exists a rounded solution $X'$ satisfying $$X' \mathbf{1}_r = r \text{ and } X'^T \mathbf{1}_c = c,$$ such that $$\left|\left<C,X'\right>-\left<C,X\right>\right|\le \epsilon \norm{C}_\infty.$$ This loss in the objective is negligible compared to the total transport cost in most real world datasets.

\subsection{Combinatorial Optimal Transport}

\subsubsection{Variations of the Hungarian algorithm.}
Combinatorial methods such as the Hungarian algorithm can be used to solve the min cost matching problem on a bipartite graph, the fastest implementation of which converges in $O(n (m + n \log n))$ time, where $n$ and $m$ are the number of vertices and edges, respectively.
\textbf{Gabow} \cite{gabow84} developed a scaling algorithm for min cost matching, 
where the costs are doubled $\log(N)$ times, here $N$ is the largest magnitude of a cost. This algorithm converges in $O(m n^{\nicefrac{3}{4}} \log(N)$ time. 

\subsubsection*{Gabow and Tarjan} \cite{GabowT91} improved this result with a different scaling algorithm, wherein the costs are doubled $\log(nN)$ times instead of $\log(N)$ times, using an additional $\log(n)$ scalings to ensure that the last approximate optimum is exact. This scaling mechanism allows the algorithm to simulate both the Hungarian algorithm and the Hopcraft-Karp algorithm simultaneously, in the low-cost regime. The increased number of scalings improves the time to convergence to $O(m \sqrt{n} \log (nN))$. This matchings algorithm can be extended to produce an optimal transport flow in $O((m \sqrt{nS} +S\log(S))\log (nN))$ time, where $S=\sum_i r_i$ is the sum over all supplies.

More recently, \textbf{Lahn \textit{et. al.}} \cite{LahnMR19} adapted the Gabow-Tarjan scaling algorithm to bound the runtime for approximate optimal transport flows with additive errors by $O(n^2 (N/\delta) + n (N/\delta)^2)$, where $\delta$ is the additive error.

\subsubsection*{Batched KM.} \label{sec-batch-km}
We develop a new variant of the
Hungarian (Kuhn-Munkres) algorithm using a single scaling factor, which 
while does not have improved theoretical performance over classical KM, outperforms classical KM in practice, as shown in Table~\ref{tab:approx_unit}.
The technique is based on the observation that up to a small error,
most of the OT data sets are well-approximated by one
where all edge costs are small integers.
The algorithm routes maximal sets of augmenting paths among
the tight edges, and augments the matching with them.
When no such augmentations are found,
the algorithm then adjusts the dual potentials across
a cut separating the left sources and the right sinks
in the same manner as KM.

The main advantage of this `batched KM'
method is that the costs
spent on finding the augmenting paths is amortized
over the total number of edges after each step,
and the number of tight edges at each round is small.
Intuitively, the small edge weights mean that
the maximum distance (dual potential) at the end is small.
So each round of maximal path augmentation
is analogous to computing a maximum matching on a
(dense) graph, which in practice is a much easier problem.

\subsubsection{MC64 and Matchbox}

\subsubsection*{MC64.}
MC64 is a matching routine originally developed by Duff and Koster \cite{mc64DuffKoster01}, and is popularly known as the MC64 subroutine in the HSL Mathematical Software Library \cite{hslLib}. MC64 produces a matrix column permutation that maximizes certain aspects of the diagonal terms, such as the smallest term or the sum or product of the diagonal terms. This can be easily adapted to solve bipartite matching problems. We benchmark using SuperLU's \cite{superlu} implementation. 

\subsubsection*{Matchbox.}
Matchbox\footnote{\url{https://github.com/CSCsw/matchbox}} is a toolkit for exact and approximate matching problems developed at Purdue University.
These algorithms are directly applicable to the assignment problem, and we benchmark using the bipartite matching algorithm that optimizes with respect to edge weights.

\subsubsection{The auction algorithm.}
The auction algorithm \cite{bertsekas89} is a combinatorial algorithm that approximately solves the min cost matching problem on a bipartite graph. It maximizes the total reward $\sum_{ij} R_{ij} X_{ij}$ given bidders $u_i$, prices $p_j$, and pairwise rewards $R_{ij}$. The auction algorithm repeatedly attempts to find a matching where each individual reward is maximized at the current $p_{j}$, and while this not possible, incrementally raises $p_j$ until such a matching is possible. This can be used to find the optimal flow by setting $R_{ij} = -C_{ij}$. 
Note that the bidding and price increment phases can be parallelized. $\epsilon$-scaling can be used to accelerate the algorithm. Here we benchmark the auction algorithm in the unit-capacitated setting with $\epsilon$-scaling.

\section{Experiments}

\subsection{Experimental Setup.}
All our code is in C++.
All experiments are single threaded and performed on an Intel(R) Xeon(R) Platinum 8168 @ 2.70GHz cpu 
with Ubuntu 18.04.2 LTS (GNU/Linux 4.18.0-1024-azure x86\_64).
For matrix operations in the implementations of Sinkhorn and Greenkhorn, we use the highly optimized OpenBLAS \cite{OpenBLAS} for Basic Linear Algebra Subprograms (BLAS).


\subsection{Datasets.}
We consider four settings to compute the OT problem: between integral points filling geometric regions in the shape of a disk and a square, between word embeddings of texts, between MNIST images, and between CIFAR images. We release an expanded version of these datasets for public benchmarking.

\subsubsection*{CircleSquare.}
CircleSquare describes a min cost matching problem between integral points. Given a square region and circular region in $\mathbb{R}^2$ that share the same center and contain the same number of integral points, CircleSquare asks for a matching between the two sets of integral points such that the total Euclidean distance between matched points is minimum. 

\subsubsection*{NLP.}
We create distributions from disjoint portions of the novel \textit{The Count of Monte Cristo}. Each sample contains 900 lines of text, with varying lengths. The text is tokenized with AllenNLP \cite{AllenNLP} and stopwords are removed. The resulting tokens are embedded into $\mathbb{R}^{100}$ with 100-dimensional GloVe \cite{glove} word embeddings, creating a $N\times 100$ matrix, where $N$ is the number of tokens in the distribution. The capacities are integral vectors of size the number of tokens, and the cost matrix contains pairwise Euclidean distances between the token embeddings in a pair of distributions.

\subsubsection*{MNIST.}
The MNIST \cite{LeCunC10} dataset contains black and white images of size $28 \times 28$ pixels, thus each image gives a capacity distribution of length-$784$ ($28^2$), where capacity values are determined by pixel intensities. Since this vector is a priori sparse, we keep only the nonzero coordinates for efficiency. To compute the optimal transport between two distributions, the cost matrix consists of the pairwise Euclidean distances between the $(x, y)$ coordinates in each distribution. The supply and demand capacity vectors are appropriately normalized and rounded such that the supply and demand sum to the same. 

\subsubsection*{CIFAR.}
The CIFAR\footnote{\url{https://www.cs.toronto.edu/~kriz/cifar.html}} \cite{Krizhevsky09} dataset contains color images of size $32 \times 32$ pixels. To construct a distribution from an image, we represent each pixel as a $(x, y, r, g, b)$ tuple, and scale each tuple such that the range of each coordinate is $[0, 1]$. Each image is thus represented as a $1024\times 5$ matrix, and the cost matrix between two images is computed as the Euclidean distance matrix between the two $1024\times 5$ representations. The capacities are integral vectors of size $1024$, appropriately normalized and rounded such that the supply and demand capacities sum to the same.

Table~\ref{tab:datasets} contains metadata about the various datasets that we test the algorithms on.

\begin{table*}[!hbt]
    \centering
    \begin{tabular}{|c|c|c|c|c|c|}\hline
    Dataset & $n$ & $m$ & supply/demand range & total demand & cost range \Bstrut\\
    \hline\textbf{CS100} & 100 & 100 & 1 - 1 & 100 & 1365 - 1382653 \Tstrut\\
\textbf{CS900} & 900 & 900 & 1 - 1 & 900 & 445 - 1554383 \\
\textbf{CS2500} & 2500 & 2500 & 1 - 1 & 2500 & 0 - 1571480\\
\textbf{CS4900} & 4900 & 4900 & 1 - 1 & 4900 &  8 - 1579922 \Bstrut\\ 
\hline
\textbf{mnist0}  & 116 & 169 & 54 - 13818 & 999929 & 0 - 234 \Tstrut\\
\textbf{mnist1}  & 165 & 172 & 144 - 9194 & 999961 & 0 - 241 \\
\textbf{mnist2}  & 64 & 136 & 218 - 25833 & 999961 & 0 - 226 \\
\textbf{mnist3}  & 193 & 168 & 68 - 8671 & 999933 & 0 - 209\\
\textbf{mnist4}  & 120 & 75 & 208 - 15741 & 999945 & 0 - 204\\
\textbf{mnist5} & 82 & 137 & 72 - 18332 & 999950 & 0 - 219\\ 
\textbf{mnist6}  & 135 & 148 & 43 - 12037 & 999926 & 0 - 219 \\
\textbf{mnist7}  & 129 & 134 & 88 - 12107 & 999948 & 0 - 262 \\
\textbf{mnist8}  & 174 & 210 & 65 - 8297 & 999920 & 0 - 220 \\
\textbf{mnist9}  & 176 & 106 & 95 - 15903 & 999942 & 0 - 233 \Bstrut\\
\hline
\textbf{NLP1} & 1389 & 1638 & 9 - 7270 & 81400 & 0 - 11577502 \Tstrut\\
\textbf{NLP2}  & 1860 & 1614 & 10 - 7020 & 89030 & 0 - 10839515   \\
\textbf{NLP3}  & 1556 & 1555 & 9 - 6760 & 80210 & 0 - 11512870 \\
\textbf{NLP4}  & 1788 & 1757 & 10 - 6390 & 88190 & 0 - 11480738 \\
\textbf{NLP5}  & 1705 & 1775 & 8 - 6729 & 81820 & 0 - 11401853 \Bstrut  \\
    
\hline

\textbf{CIFAR1} & 1024 & 1024 & 1 - 1 & 1024 & 5 - 610 \Tstrut\\
\textbf{CIFAR2} & 1024 & 1024 & 1 - 1 & 1024 & 9 - 611\\
\textbf{CIFAR3} & 1024 & 1024 & 1 - 1 & 1024 & 10 - 619 \\
\textbf{CIFAR4} & 1024 & 1024 & 1 - 1 & 1024 & 5 - 568 \\
\textbf{CIFAR5} & 1024 & 1024 & 1 - 1 & 1024 & 3 - 600\\
\textbf{CIFAR6} & 1024 & 1024 & 1 - 1 & 1024 & 5 - 654 \Bstrut\\

\hline
\end{tabular}
    \vspace{2mm}
    \caption{Details of datasets used in our experiments.}
    \label{tab:datasets}
\end{table*}

\section{Results}

\subsection{Algorithm configurations.}


The following is a brief description of the parameters and settings used in the implementations of the algorithms described in Section~\ref{sec:algos}.

\subsubsection*{Lemon NS.}
This is an exact solution using the network simplex algorithm.
This is implemented in C++ by using the network simplex module from Lemon\footnote{\url{https://lemon.cs.elte.hu/trac/lemon}} \cite{DezsJK11}
as a blackbox which computes minimum cost flows.

\subsubsection*{KM and variants.}
We use an implementation using the Kuhn-Munkres (Hungarian) algorithm
given by Ivanov Maxim\footnote{\url{http://e-maxx.ru/algo/assignment_hungary}}, which is an exact algorithm.

We present the results of exact KM and our fast variant \textbf{batched KM} in Table~\ref{tab:exact_unit} and Table~\ref{tab:approx_unit}.

\subsubsection*{MC64 and Matchbox.}

We use SuperLU's implementation \footnote{\url{https://github.com/xiaoyeli/superlu}} for MC64 and the original implementation \footnote{\url{https://github.com/CSCsw/matchbox}} for Matchbox. These implementations are applicable to graphs with unit-capacitated supplies and demands.

\subsubsection*{Auction.}
In the case when the edge weights are integral, the auction algorithm gives an exact solution when the slack parameter, $\epsilon$, is less than $1/n$, since a feasible solution produced by the auction algorithm is guaranteed to be within $n\epsilon$ of the optimal cost. In practice, however, increasing the value of $\epsilon$ can continue to yield exact solutions since integral edge weights lead to discrete solutions. We use the highest $\epsilon$ that still gives
an exact solution, to obtain the fastest possible runtime.

However, the relation between runtimes
and $\epsilon$ is not necessarily monotonic, especially
when $\epsilon \gg 1$.
In our implementation, we allow for some slack in the demand being satisfied by the algorithm, in particular, we make sure all but two nodes are matched.
For the remaining two nodes on either side,
we simply pick the cheaper assignment among the two possibilites.

$\epsilon$ scaling is often part of the auction algorithm to speed up convergence. Here $\epsilon$ is regularly decreased to allow for more exact routing of flows of similar costs.
We refer to this version of the algorithm as \textbf{AuctionScaled}.

\subsubsection*{Sinkhorn.}
The approximation factor and runtime of the Sinkhorn algorithm
depend on our choice of $\eta$ (the regularization parameter)
and $\epsilon$ (the maximum allowed distance to feasible solution).

We first fix $\epsilon$ to be $0.1\%$ of the total demand (or equivalently,
total supply). In most popular implementations of Sinkhorn,
this leftover demand is satsified using a rounding technique,
as discussed in Section~\ref{sec:algos}. However, rounding results in a significant timing overhead without always bringing the objective closer to optimality (in fact as discussed in section~\ref{sec-rounding}, rounding often results in a worse OT approximation in datasets tested), our implementations do not round.

There is a trade-off between the regularization parameter $\eta$ and the
approximation factor: picking higher values of $\eta$ leads to better approximations but longer runtimes.
Therefore in our experiments, we pick the smallest possible $\eta$ for which the algorithm
converges to a solution which is 110\% of the optimal routing cost
(computed by the exact algorithm network simplex).



\subsubsection*{Greenkhorn.}
Similar to Sinkhorn, we fix $\epsilon$ to be $0.1\%$ of the total demand, and pick the smallest possible regularization parameter $\eta$ for which we obtain a 1.1-approximate solution.
Note that the relation between the approximation factor and the regularization
parameter $\eta$ is not affected by the order in which the rows and columns are updated.
Hence, the values of $\eta$ from Sinkhorn experiments are also applicable for Greenkhorn.

To maintain parity with Sinkhorn, we only compute the total flow cost
once every $(m+n)/2$ iterations of Greenkhorn - corresponding to one iteration of Sinkhorn on average.



\subsection{Comparison between algorithms.}

We first compare the algorithms on graphs with unit demands and supplies, i.e.,
algorithms for the assignment problem.
In Table~\ref{tab:exact_unit}, we compare the runtimes of exact algorithms
for the problem.

\begin{table*}[!ht]
    \centering
    \begin{tabular}{|c|c|c|c|c|c|}\hline
{Graph} &  Lemon NS &  KM & Auction &  Matchbox & MC64 \Tstrut\Bstrut\\
\hline
\textbf{CS100}  & 0.001s & 0.001s & $<$0.001s & 0.001s & 0.001s \Tstrut\\
\textbf{CS900}  & 0.076s & 0.155s &    0.405s & 0.134s  & 0.157s \\
\textbf{CS2500} & 1.253s & 1.469s &    5.027s & 2.692s & 2.368s  \\
\textbf{CS4900} & 7.411s & 14.53s & - & 25.06s & 23.02s \Bstrut\\
\hline
\textbf{CIFAR1} & 0.178s & 0.368s & 0.055s & 0.346s & 0.359s \Tstrut\\
\textbf{CIFAR2} & 0.151s & 0.142s & 0.073s & 0.212s & 0.206s  \\
\textbf{CIFAR3} & 0.167s & 0.297s & 0.071s & 0.379s & 0.536s  \\
\textbf{CIFAR4} & 0.178s & 0.297s & 0.133s & 0.351s & 0.219s  \\
\textbf{CIFAR5} & 0.216s & 0.288s & 0.172s & 0.308s & 0.354s  \\
\textbf{CIFAR6} & 0.192s & 0.528s & 0.351s & 0.605s & 0.678s \Bstrut\\
\hline
\end{tabular}
    \vspace{2mm}
    \caption{Comparison of runtimes between various methods for \textbf{exact} OT computation on graphs with \textit{unit} demands and supplies (assignment problem).}
    \label{tab:exact_unit}
\end{table*}

\begin{figure}[!htb]
  \centering
  \includegraphics[width=\linewidth]{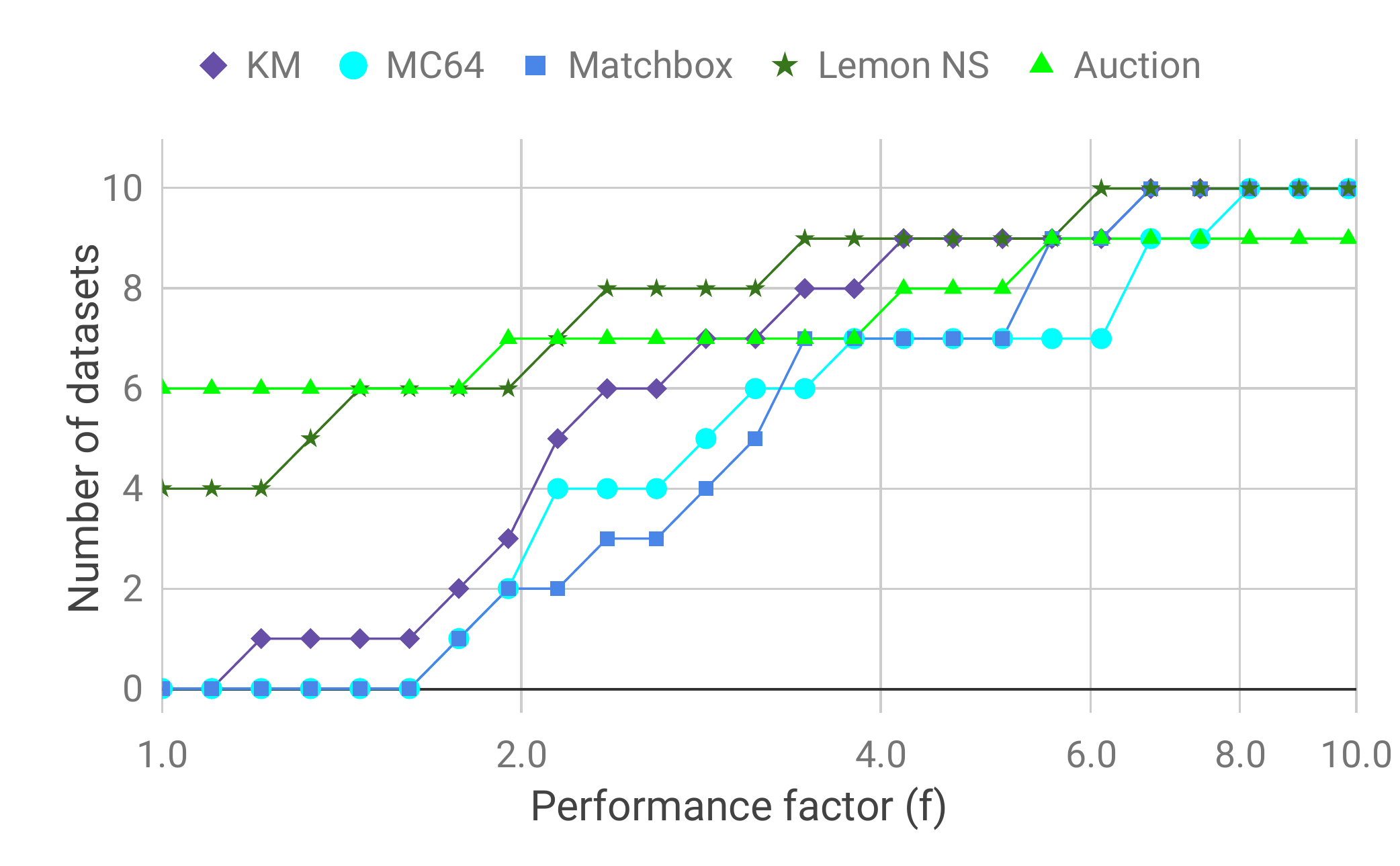}  
  \caption{Performance profile of exact algorithms for OT.}
  \label{fig:perf1}
\end{figure}

At a high level, while the gap between algorithms is not large, we observe that Lemon NS outperforms
other algorithms in most datasets.
Note that Auction times out before achieving a perfect matching on CS4900.
We visualize this in the following performance profile \cite{Dolan2002} of the algorithms in Figure~\ref{fig:perf1}.
For an algorithm, the \textit{performance factor} $f \in [1,\infty)$ is defined as the factor
by which it is slower than the fastest algorithm on a particular dataset.
The fastest algorithm has $f=1$.
For an algorithm, we plot $f$ versus the number of datasets for which 
the algorithm has a performance factor at most $f$.

Next, we compare the approximation algorithms for the assignment
problem, shown in Table~\ref{tab:approx_unit}. For fair comparison, we fix the desired
approximation factor to be 1.1 across all algorithms. The sparse batched KM variant introduced in Section~\ref{sec-batch-km} significantly outperforms the other methods profiled. 

\begin{table*}[!ht]
    \centering
    \begin{tabular}{|c|c|c|c|c|c|}\hline
{Graph} &  Batched KM &  Sinkhorn & Greenkhorn & Auction & AuctionScaled\Tstrut\Bstrut\\
\hline
\textbf{CS100}  & $<$0.001s & 0.006s & 0.402s & $<$0.001s & $<$0.001s \Tstrut\\
\textbf{CS900}  &    0.014s & 3.225s & 28.03s &    0.562s &    0.371s  \\
\textbf{CS2500} &    0.332s & 127.8s & 650.8s &    11.75s &    8.592s \\
\textbf{CS4900} &    1.350s & 620.1s & 3053s  &    119.6s &    49.51s \Bstrut\\
\hline
\textbf{CIFAR1} & 0.010s & 0.86s  & 16.81s & 0.453s &  0.166s \Tstrut\\
\textbf{CIFAR2} & 0.008s & 0.99s  & 18.75s & 0.141s &  0.116s \\
\textbf{CIFAR3} & 0.008s & 0.51s  & 11.03s & 0.126s &  0.053s \\
\textbf{CIFAR4} & 0.008s & 0.49s  & 8.865s & 0.297s &  0.154s \\
\textbf{CIFAR5} & 0.011s & 0.36s  & 5.394s & 0.047s &  0.031s \\
\textbf{CIFAR6} & 0.010s & 0.26s  & 6.078s & 0.218s &  0.078s \Bstrut\\
\hline
\end{tabular}
    \vspace{2mm}
    \caption{Comparison of runtimes between various \textbf{approximation} algorithms for OT on graphs with \textit{unit} demands and supplies (assignment problem). Here, all methods produce a 1.1-approximate solution.}
    \label{tab:approx_unit}
\end{table*}


\begin{figure}[!htb]
  \centering
  \includegraphics[width=\linewidth]{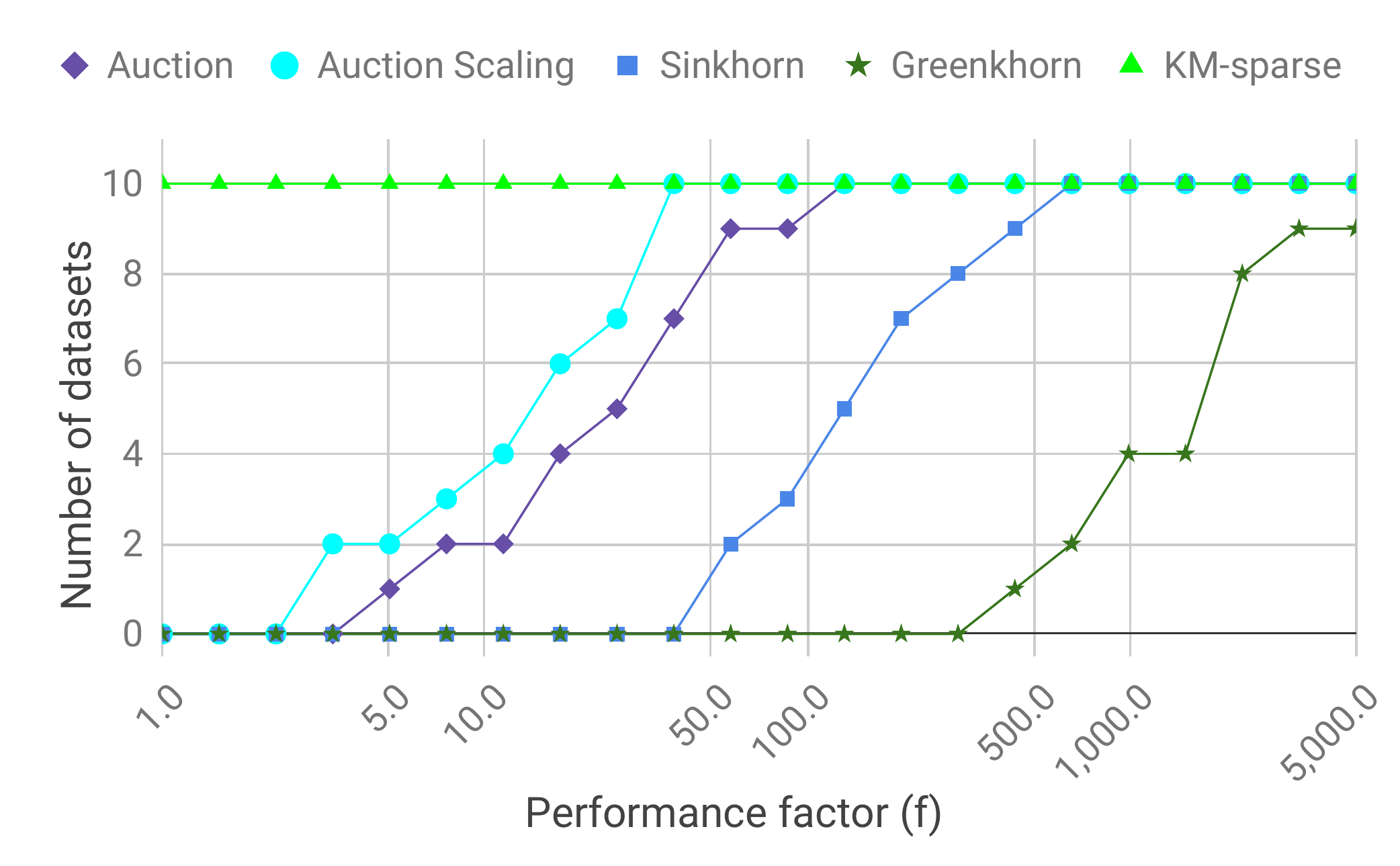}  
  \caption{Performance profile of approximation algorithms for OT.}
  \label{fig:perf2}
\end{figure}

Lastly, we evaluate algorithms which also allow for non-unit demands and supplies.
Among the algorithms we discuss, only Lemon NS, Sinkhorn and Greenkhorn
satisfy this property.
We compare their runtimes in Table~\ref{tab:runtimes}

\begin{table}[!ht]
    \centering
    \begin{tabular}{|c|c|c|c|}\hline
{Graph} &  Lemon NS & Sinkhorn & Greenkhorn \Tstrut \\
{} & (exact) & (1.1-approx) & (1.1-approx)\Bstrut\\
\hline

\textbf{mnist0} & 0.002s & 0.004s & 0.071s  \Tstrut\\
\textbf{mnist1} & 0.003s & 0.008s & 0.117s \\
\textbf{mnist2} & 0.001s & 0.001s & 0.017s  \\
\textbf{mnist3} & 0.003s & 0.025s & 0.450s \\
\textbf{mnist4} & 0.001s & 0.001s & 0.030s  \\
\textbf{mnist5} & 0.001s & 0.001s & 0.016s  \\
\textbf{mnist6} & 0.002s & 0.008s & 0.129s  \\
\textbf{mnist7} & 0.002s & 0.003s & 0.035s  \\
\textbf{mnist8} & 0.003s & 0.005s & 0.091s \\
\textbf{mnist9} & 0.002s & 0.005s & 0.079s \Bstrut\\
\hline
\textbf{NLP1}   & 0.454s & 1.950s & 9.246s \Tstrut\\
\textbf{NLP2}   & 0.582s & 2.901s & 9.813s \\
\textbf{NLP3}   & 0.464s & 2.499s & 7.838s \\
\textbf{NLP4}   & 0.558s & 1.776s & 9.252s \\
\textbf{NLP5}   & 0.547s & 2.519s & 9.801s \\
\hline
\end{tabular}
    \vspace{2mm}
    \caption{Comparison of runtimes between various methods for OT on graphs with \textit{non-unit} demands and supplies.}
    \label{tab:runtimes}
\end{table}

From Table~\ref{tab:runtimes},
we see that despite allowing a 10\% margin of error to Sinkhorn and Greenkhorn,
they do not outperform Lemon NS on any dataset.
In fact, Greenkhorn is at least an order of magnitude slower.

\subsection{Effect of regularization in matrix scaling techniques.}

In both matrix scaling methods - Greenkhorn and Sinkhorn,
the accuracy of the algorithm depends crucially on the choice
of $\eta$. As we increase the value of $\eta$,
the algorithm requires more iterations to converge and naturally
also approaches a better approximation factor.

\begin{figure}[!ht]
\begin{subfigure}{0.5\textwidth}
  \centering
  \includegraphics[width=\linewidth]{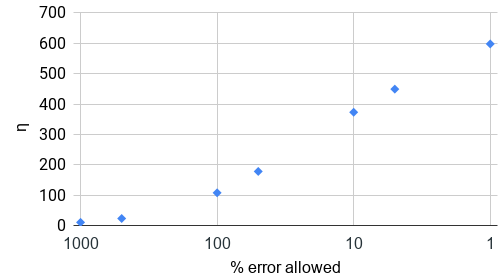}  
  \caption{CS100}
  \label{fig:err1}
\end{subfigure}
\begin{subfigure}{0.5\textwidth}
  \centering
  \includegraphics[width=\linewidth]{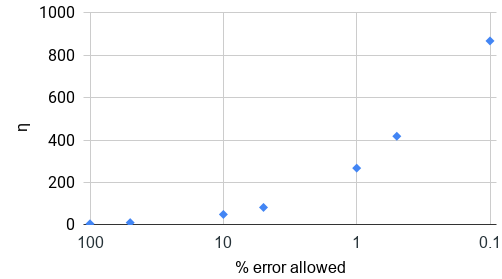}  
  \caption{mnist0}
  \label{fig:err2}
\end{subfigure}
\caption{Trade-off between regularization parameter $\eta$ and accuracy in Sinkhorn and Greenkhorn.}
\label{fig:accuracy}
\end{figure}

Figure~\ref{fig:accuracy} shows the relation between the accuracy of
matrix scaling methods with increasing values of $\eta$.
Note that this value of $\eta$ is selected assuming that
the maximum value of the cost matrix is 1.
This allows for a fair comparison between different datasets.
The matrices have been scaled appropriately to accommodate this.

As can be seen from Figure~\ref{fig:accuracy},
significantly high values of the regularization parameter $\eta$ are required
for the algorithms to reach an approximation arbitrarily close to optimal.
This creates many issues in practice because higher values
of $\eta$ require higher float precision from the compiler (due to very small
exponentials appearing).

\begin{figure}[!ht]
\begin{subfigure}{0.5\textwidth}
  \centering
  \includegraphics[width=\linewidth]{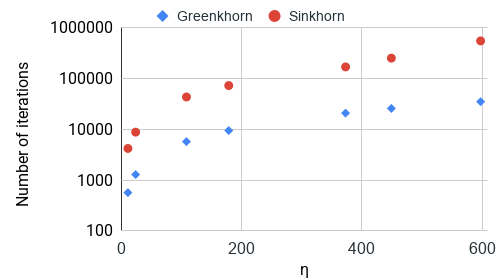}  
  \caption{CS100}
  \label{fig:conv1}
\end{subfigure}
\begin{subfigure}{0.5\textwidth}
  \centering
  \includegraphics[width=\linewidth]{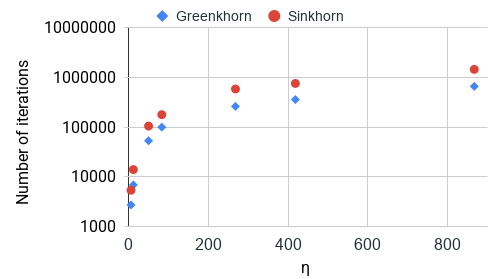}  
  \caption{mnist0}
  \label{fig:conv2}
\end{subfigure}
\caption{Convergence analysis of Sinkhorn and Greenkhorn}
\label{fig:convergence}
\end{figure}

Figure~\ref{fig:convergence} shows the number of iterations needed for the algorithm
to converge as a function of the desired approximation factor (or equivalently, $\eta$).

We see that the number of iterations grows super-linearly,
and this can cause algorithms to be very slow when trying to reach a solution with high accuracy.

\section{Conclusion}

This work systematically profiled the performance of two broad classes of algorithms to solve the optimal transport problem: combinatorial and matrix-scaling based methods. We observed that despite the design of asymptotically faster algorithms for optimal transport, combinatorial algorithms such as network simplex, auction algorithms, and adaptations of Kuhn-Munkres are able to significantly outperform newer matrix rescaling based algorithms in speed, despite a 10\% error allowance for the latter. Matrix scaling methods still enjoy the advantage of being differentiable -- hence more easily adaptable to neural network learning -- as opposed to the combinatorial methods studied, this work leaves the door open for developing differentiable analogs of the combinatorial algorithms. 





\begin{thebibliography}{99}

\bibitem{AhujaMO93}
Ravindra~K. Ahuja, Thomas~L. Magnanti, and James~B. Orlin.
\newblock {\em Network flows - theory, algorithms and applications}.
\newblock Prentice Hall, 1993.

\bibitem{AltschulerWR17}
Jason Altschuler, Jonathan Weed, and Philippe Rigollet.
\newblock Near-linear time approximation algorithms for optimal transport via
  sinkhorn iteration.
\newblock In {\em Advances in Neural Information Processing Systems 30: Annual
  Conference on Neural Information Processing Systems 2017, 4-9 December 2017,
  Long Beach, CA, {USA}}, pages 1964--1974, 2017.

\bibitem{ArjovskyCB17}
Mart{\'{\i}}n Arjovsky, Soumith Chintala, and L{\'{e}}on Bottou.
\newblock Wasserstein generative adversarial networks.
\newblock In {\em Proceedings of the 34th International Conference on Machine
  Learning, {ICML} 2017, Sydney, NSW, Australia, 6-11 August 2017}, pages
  214--223, 2017.

\bibitem{BenamouCCNP15}
Jean{-}David Benamou, Guillaume Carlier, Marco Cuturi, Luca Nenna, and Gabriel
  Peyr{\'{e}}.
\newblock Iterative bregman projections for regularized transportation
  problems.
\newblock {\em {SIAM} J. Scientific Computing}, 37(2), 2015.

\bibitem{bertsekas89}
Dimitri Bertsekas and David Castanon.
\newblock The auction algorithm for the transportation problem.
\newblock {\em Annals of Operations Research}, 20:67--96, 1989.

\bibitem{BigotGKL17}
J.~{Bigot}, R.~{Gouet}, T.~{Klein}, and A.~{L{\'o}pez}.
\newblock {Geodesic PCA in the Wasserstein space by convex PCA}.
\newblock {\em Annales de l'Institut Henri Poincar{\'e}, Probabilit{\'e}s et
  Statistiques, vol.~53, issue 1, pp.~1-26}, 53:1--26, February 2017.

\bibitem{BlanchetJKS18}
Jose~H. Blanchet, Arun Jambulapati, Carson Kent, and Aaron Sidford.
\newblock Towards optimal running times for optimal transport.
\newblock {\em CoRR}, abs/1810.07717, 2018.

\bibitem{Cuturi13}
Marco Cuturi.
\newblock Sinkhorn distances: Lightspeed computation of optimal transport.
\newblock In {\em Advances in Neural Information Processing Systems 26: 27th
  Annual Conference on Neural Information Processing Systems 2013. Proceedings
  of a meeting held December 5-8, 2013, Lake Tahoe, Nevada, United States.},
  pages 2292--2300, 2013.

\bibitem{DezsJK11}
Bal\'{a}zs Dezs, Alp\'{a}r J\"{u}ttner, and P{\'e}ter Kov\'{a}cs.
\newblock Lemon - an open source c++ graph template library.
\newblock {\em Electron. Notes Theor. Comput. Sci.}, 264(5):23--45, July 2011.

\bibitem{DvurechenskyGK18}
Pavel~E. Dvurechensky, Alexander Gasnikov, and Alexey Kroshnin.
\newblock Computational optimal transport: Complexity by accelerated gradient
  descent is better than by sinkhorn's algorithm.
\newblock In {\em Proceedings of the 35th International Conference on Machine
  Learning, {ICML} 2018, Stockholmsm{\"{a}}ssan, Stockholm, Sweden, July 10-15,
  2018}, pages 1366--1375, 2018.

\bibitem{gabow84}
Harold Gabow.
\newblock Scaling algorithms for network problems.
\newblock {\em Journal Of Computer AND System Sciences}, 31, 1985.

\bibitem{GabowT91}
Harold~N. Gabow and Robert~E. Tarjan.
\newblock Faster scaling algorithms for general graph matching problems.
\newblock {\em J. ACM}, 38(4):815--853, October 1991.

\bibitem{hslLib}
HSL (formerly the Harwell Subroutine Library)
\newblock HSL, A Collection of Fortran Codes for Large Scale Scientific Computation.
\newblock {\em http://www.hsl.rl.ac.uk}.

\bibitem{AllenNLP}
Matt Gardner, Joel Grus, Mark Neumann, Oyvind Tafjord, Pradeep Dasigi,
  Nelson~F. Liu, Matthew~E. Peters, Michael Schmitz, and Luke Zettlemoyer.
\newblock Allennlp: {A} deep semantic natural language processing platform.
\newblock {\em CoRR}, abs/1803.07640, 2018.

\bibitem{GenevayCPB16}
Aude Genevay, Marco Cuturi, Gabriel Peyr{\'{e}}, and Francis~R. Bach.
\newblock Stochastic optimization for large-scale optimal transport.
\newblock In {\em Advances in Neural Information Processing Systems 29: Annual
  Conference on Neural Information Processing Systems 2016, December 5-10,
  2016, Barcelona, Spain}, pages 3432--3440, 2016.

\bibitem{HoNYBHP17}
Nhat Ho, XuanLong Nguyen, Mikhail Yurochkin, Hung~Hai Bui, Viet Huynh, and
  Dinh~Q. Phung.
\newblock Multilevel clustering via wasserstein means.
\newblock In {\em Proceedings of the 34th International Conference on Machine
  Learning, {ICML} 2017, Sydney, NSW, Australia, 6-11 August 2017}, pages
  1501--1509, 2017.

\bibitem{KolouriPTSR17}
Soheil Kolouri, Se~Rim Park, Matthew Thorpe, Dejan Slepcev, and Gustavo~K.
  Rohde.
\newblock Optimal mass transport: Signal processing and machine-learning
  applications.
\newblock {\em {IEEE} Signal Process. Mag.}, 34(4):43--59, 2017.

\bibitem{Krizhevsky09}
Alex Krizhevsky.
\newblock Learning multiple layers of features from tiny images.
\newblock Technical report, Canadian Institute for Advanced Research, 2009.

\bibitem{Kuhn55}
H.~W. Kuhn.
\newblock The hungarian method for the assignment problem.
\newblock {\em Naval Research Logistics Quarterly}, 2(1‐2):83--97, 1955.

\bibitem{Kuhn56}
H.~W. Kuhn.
\newblock {Variants of the hungarian method for assignment problems}.
\newblock {\em Naval Research Logistics Quarterly}, 3(4):253--258, December
  1956.

\bibitem{KusnerSKW15}
Matt~J. Kusner, Yu~Sun, Nicholas~I. Kolkin, and Kilian~Q. Weinberger.
\newblock From word embeddings to document distances.
\newblock In {\em Proceedings of the 32nd International Conference on Machine
  Learning, {ICML} 2015, Lille, France, 6-11 July 2015}, pages 957--966, 2015.

\bibitem{LahnMR19}
Nathaniel Lahn, Deepika Mulchandani, and Sharath Raghvendra.
\newblock A graph theoretic additive approximation of optimal transport.
\newblock {\em CoRR}, abs/1905.11830, 2019.

\bibitem{LeCunC10}
Yann LeCun and Corinna Cortes.
\newblock {MNIST} handwritten digit database.
\newblock http://yann.lecun.com/exdb/mnist/, 2010.

\bibitem{LeeS14}
Yin~Tat Lee and Aaron Sidford.
\newblock Path finding methods for linear programming: Solving linear programs
  in {\~{o}}(vrank) iterations and faster algorithms for maximum flow.
\newblock In {\em 55th {IEEE} Annual Symposium on Foundations of Computer
  Science, {FOCS} 2014, Philadelphia, PA, USA, October 18-21, 2014}, pages
  424--433, 2014.

\bibitem{LinHJ19}
Tianyi Lin, Nhat Ho, and Michael~I. Jordan.
\newblock On efficient optimal transport: An analysis of greedy and accelerated
  mirror descent algorithms.
\newblock In {\em Proceedings of the 36th International Conference on Machine
  Learning, {ICML} 2019, 9-15 June 2019, Long Beach, California, {USA}}, pages
  3982--3991, 2019.

\bibitem{mc64DuffKoster01}
Iain Duff and Jacko Koster.
\newblock On algorithms for permuting large entries to the diagonal of a sparse matrix.
\newblock In {\em SIAM Journal on Matrix Analysis and Applications}, 22:973.996, 2001.

\bibitem{superlu}
X. Sherry Li, Jim Demmel, John Gilbert, Laura Grigori, Yang Liu, Piyush Sao, Meiyue Shao, and Ichitaro Yamazaki.
\newblock  Supernodal LU.
\newblock {https://portal.nersc.gov/project/sparse/superlu}.

\bibitem{Munkres57}
J.~Munkres.
\newblock Algorithms for the assignment and transportation problems.
\newblock {\em Journal of the Society for Industrial and Applied Mathematics},
  5(1):32--38, 1957.

\bibitem{glove}
Jeffrey Pennington, Richard Socher, and Christopher~D. Manning.
\newblock Glove: Global vectors for word representation.
\newblock In {\em Empirical Methods in Natural Language Processing (EMNLP)},
  pages 1532--1543, 2014.

\bibitem{Quanrud19}
Kent Quanrud.
\newblock Approximating optimal transport with linear programs.
\newblock In {\em 2nd Symposium on Simplicity in Algorithms, SOSA@SODA 2019,
  January 8-9, 2019 - San Diego, CA, {USA}}, pages 6:1--6:9, 2019.

\bibitem{RubnerTG00}
Yossi Rubner, Carlo Tomasi, and Leonidas~J. Guibas.
\newblock The earth mover's distance as a metric for image retrieval.
\newblock {\em International Journal of Computer Vision}, 40(2):99--121, 2000.

\bibitem{SandlerL11}
Roman Sandler and Michael Lindenbaum.
\newblock Nonnegative matrix factorization with earth mover's distance metric
  for image analysis.
\newblock {\em {IEEE} Trans. Pattern Anal. Mach. Intell.}, 33(8):1590--1602,
  2011.

\bibitem{SolomonRGB14}
Justin Solomon, Raif~M. Rustamov, Leonidas~J. Guibas, and Adrian Butscher.
\newblock Wasserstein propagation for semi-supervised learning.
\newblock In {\em Proceedings of the 31th International Conference on Machine
  Learning, {ICML} 2014, Beijing, China, 21-26 June 2014}, pages 306--314,
  2014.

\bibitem{OpenBLAS}
Zhang Xianyi, Martin Kroeker, Werner Saar, Wang Qian, Zaheer Chothia, Chen
  Shaohu, and Luo Wen.
\newblock Openblas: An optimized blas library.
\newblock \url{https://www.openblas.net/}, 2019.

\bibitem{Dolan2002}
Elizabeth D. Dolan, and Jorge J. Mor{\'e}.
\newblock Benchmarking optimization software with performance profiles.
\newblock Mathematical Programming, 91 (2002), pp. 201-213
\end{thebibliography}
\end{document}